\documentclass{sig-alternate}

\begin{document}
%
\conferenceinfo{DISIO 2010}{March 15, Torremolinos, Malaga, Spain.}
\CopyrightYear{2010} 
\crdata{78-963-9799-87-5}  

\title{Living City, a Collaborative Browser-based Massively Multiplayer Online Game 
}

%
%
%
%
%

\numberofauthors{3} 

%
\author{
  \alignauthor Emilio Ferrara\\ \affaddr Dept. of Mathematics\\ \affaddr University of Messina \email{emilio.ferrara@unime.it}
  \alignauthor Giacomo Fiumara\\ \affaddr Dept. of Physics\\ \affaddr University of Messina \email{giacomo.fiumara@unime.it}
  \alignauthor Francesco Pagano\\ \affaddr DSI\\ \affaddr University of Milan \email{francesco.pagano@unimi.it}}
\date{}
    
\maketitle

\begin{abstract}
This work presents the design and implementation of our Browser-based Massively Multiplayer Online Game, \textit{Living City}, a simulation game fully developed at the University of Messina.
\textit{Living City} is a persistent and real-time digital world, running in the Web browser environment and accessible from users without any client-side installation. 

Today \textit{Massively Multiplayer Online Games} attract the attention of Computer Scientists both for their architectural peculiarity and the close interconnection with the social network phenomenon. 

We will cover these two aspects paying particular attention to some aspects of the project: game balancing (e.g. algorithms behind time and money balancing); business logic (e.g., handling concurrency, cheating avoidance and availability) and, finally, social and psychological aspects involved in the collaboration of players, analyzing their activities and interconnections.

\end{abstract}



\\
\section{Introduction}
Multiplayer Online Games (MOGs) are by now a market and a social phenomenon. They are also capturing the attention both of academia and industry because of the many common problems, related to their distributed nature, which can be approached and solved using enterprise solutions and novel techniques.

MOGs can be classified in several ways, for instance considering the system environment, the number of players, by the action being in real-time or by turns, or just by game genres. 
Actually, the most attractive MOGs family is the one called \textit{Massively Multiplayer Online Games} (MMOGs), which differs from other MOGs in that they are played over the Internet by thousands of players interacting in the same persistent real-time world. 

Depending on the game genre, we classify MMOGs in categories, i.e. MMORPGs (Role Play), MMOFPS (First-Person Shooter), MMOSGs (Strategic), and so on. 
Depending on the game platform we have another taxonomy, classifying MMMOGs as playable on mobile devices, and BBMMOGs (Browser based), differing from standard ones, e.g. games for PC or console, because they are accessible from the World Wide Web without any client-side installation.

The aim of this work is to introduce \textit{Living City}, a BBMMOG we designed and implemented at the University of Messina, focusing on some interesting aspects that we encountered during these steps and analyzing results obtained after almost a year of public testing. 
The first \textit{Living City} \footnote{http://www.livingcity.it} public beta version has been played by almost ten thousand users registered during this period, providing us a good amount of data useful to extract relevant information about users activity and practices, especially on the social interaction among them, and tips on game balancing and optimizing the user interface.

The paper is organized as follows: in Section 2 we consider the related work on MMOGs, in particular regarding Browser-based games. 
Sections 3 and 4 cover the design and the development of the \textit{Living City} project, detailing some interesting aspects of the architecture and the business logic of the game. 
Section 5 presents an in-depth analysis on the user activity and some final considerations of social interactions among players. 

\section{Related Work}
The fast growth and the attention MMOGs captured is witnessed by several valid works on the subject; they focus on architectural issues (e.g distribution techniques \cite{Yamamoto:distributed}, load balancing \cite{Lu:Load}, persistence \cite{Zhang:Persistence}, etc.) and Software Engineering, e.g. usability \cite{Cornett:usability}, performances measurement \cite{Kim:Traffic}, services platform \cite{Shaikh:service}.

In 2005 Yan and Randell \cite{Yan:cheatingclassification} provided a useful taxonomy of cheating in online games, covering fifteen cheating techniques and approaches and their relative  countermeasures. 
A year later a remarkable game suggestion from Almeida and Gomes \cite{Almeida:advancedstrategic} introduced some interesting ideas on the BBMMOGs development. 
Niso \cite{Niso:Production} provided an exhaustive report of the production life cycle of a BBMMOG. 
In the same year Ferrara \cite{Ferrara:Progettazione}, in his M.Sc. dissertation, started the design of \textit{Living City}. 
Our ultimate goal then is to give our contribution to cover the gap between academic research and game industry in the design and implementation of \textit{Living City}.

\section{Game Design}

\subsection{Focus}
The primary idea was to focus the game concept on the players collaboration, providing motivations to cooperate in order to reach personal in-game objectives with the support of the community of other players, or, in fact, making harder or impossible to complete some tasks without cooperation and collaboration. 
This because we analyzed the actual game design trends, pointing out that, despite the fact that many commercial BBMMOGs appear to be collaborative, actually almost each gameplay discourages cooperation among players, focusing on competitive aspects of the game and relying on the game addiction. 

Common BBMMOGs genres are Sports, Role Playing, Strategy and Simulation.  
For each of these genres there are dozens of examples, and also hybridization were developed, for instance games with both Role Play and Strategic features, etc. 
Furthermore, Sports BBMMOGs generally include both management and simulation aspects. 
These three segments in most cases imply player competition.
Actually the last genre, simulation, allows both collaborative and competitive game styles. 
We chose the simulation segment to start our analysis, designing an innovative game concept.

\subsubsection{An Overview on BBMMOGs Features}
BBMMOGs can be classified according to graphical features, e.g. animated graphics, graphical user interface, etc.
Some of them are usually based on Adobe Flash \footnote{http://www.adobe.com} or Java \footnote{http://sun.java.com}, and are similar to off-the-shelf MMOGs. Others are based on HTML pages, being developed like usual Web applications. 
The former category of games implies a huge amount of work on graphical compartment and scripting, the latter requires a better design, in particular on database and interface.
Both show common advantages:

\begin{itemize}
  \item Availability
  \item Business Model
  \item Accessibility
  \item Architectural Requirements
  \item Performances
  \item Limited Development Efforts
\end{itemize}

The \textit{availability} represents a great incentive for players because they have a huge number of games, almost freely playable, and the freedom of choosing the most suitable for their expectations: indeed, at difference with common off-the-shelf games, BBMMOGs are free-of-charge, except for some features, usually presented as premium ones, which sometimes give a couple of advantages in the game to paying players, and/or are represented by special items with some singular powers. 
In these games the \textit{business model} is also built around the advertising, which often becomes the greater economic income for developers. 
Another advantage is that the game itself can be advertised on the same advertising channels, specifically oriented to the game market, so applying strategies to attract the maximum possible number of paying players.

Regarding architectural aspects, there are a couple of advantages both for players and developers, related to the Web nature of the product: the \textit{accessibility} is global because BBMMOGs cut off the distribution channels of physical products, completely relying on Internet websites; this is usually an advantage also in comparison with off-the-shelf MMOGs, which are usually sold in shops, like any other stand-alone game. 
Consequently, \textit{architectural requirements} are really low, indeed any hardware configuration that can host a Web application server fits good the work. 
A quantitative and qualitative dimension of the server configuration, related to expected traffic and quality of service we must ensure, should be designed and provided by developers: \textit{performances} will be commensurate to the strength of this architecture, but, equal investment, are much more effective and solid w.r.t. other architectures, designed with the same budget, to run standard MMOGs. 
Last but not least, the limited effort to develop BBMMOGs represents a good reason to study this field of application in academia, and represents a huge advantage, especially in industry, enabling small teams to start-up in a market segment, so rapidly developing appreciated products.

\subsubsection{Game concept}
\textit{Living City} is a simulation game designed to be collaborative and socialization oriented: the main aim for players is to manage and administrate a virtual city, embodying the role of the Mayor, including staff, economic, construction and investment management tasks,  services administration, business decision-making, etc. 
The game concept is widely inspired by the 'SimCity' computer games series, created and first developed by Will Wright in 1989, with major changes regarding the game orientation to multiplayer goals: \textit{Living City} is designed to make possible to manage collaborative tasks like services providing, building permits, licenses releasing, procurements proclaiming, etc. 
All these aspects of the game require cooperation among players and cannot be completed playing alone.
Notwithstanding it is possible to slowly advance in the game also taking not care of some of these points. 
Cooperation is a key evaluation criterion for algorithms deciding which cities should grow faster than others, which should acquire more relevance, etc.

\subsection{Structure of the Game}
\textit{Living City} is structured in the following macro areas:

\begin{itemize}
  \item Mayor and Staff;
  \item City and Buildings, and
  \item Management.
\end{itemize}

There are also additional, although not strictly game-related, areas regarding \textit{community of players}, \textit{in-game tools}, and \textit{game guide}. 
We will cover all these aspects of design and development of the simulation.

\subsubsection{Mayor and Staff}
In this first macro area, the player can access different sections to manage and monitor aspects regarding his game profile and, moreover, statistics, performances and productivity of own staff. 
A priority, in the starting phase of the game, is \textit{recruitment}: the Mayor has to cover some relevant positions (e.g. the Deputy Mayor, Public Relationships, Human Resources, Technicals and Engineers, Administrators, etc.) employing \textit{NPCs} (Non-Player Characters); their activity is controlled by the system and is fundamental for completing some tasks (e.g. managing the provision of services, issuing calls for procurements, etc.), but there are limitations on recruitment (e.g. a maximum defined number of employees in each role, considering hiring/firing effects, etc.). 
The right choices will imply better staff performances.
players have tools (e.g. each NPC has a \textit{fictitious blog} that automatically and dynamically reports his/her thoughts about the assigned tasks, a page with statistics and graphs to analyze performances of employees, etc.) and indicators (e.g. a \textit{rating scale} for public parameters) to monitor the working group.

There are also several \textit{activity configurations} to tune.
Some of them help the player to tune attitudes and behaviors in his/her staff even outside the working hours, and, most importantly, working tasks they have to complete, depending on their position. 
All these variables and results are calculated through public and hidden  parameters related to employees.
Each NPC has got common attributes (i.e. willingness to work, ability to work in team, concentration, intuition, motivation, etc.), and some expertise, related to his role: better employees cost much to the administration but ensure better performances and productivity. 

Players have to balance requirements with costs and the hidden optimization function is dynamically generated by the system, considering more than 50 variables and parameters related to the actual condition of the city, degree of development, needs, etc: this ensures that there is no simple solution to the problem of the recruitment or a standard setup of the staff which fits well in any situation. 
One of the game challenges is to find a good staff configuration over the time.

\subsubsection{City and Buildings}
The second macro-area allows players to develop structural aspects of the city: after the registration each user receives a virtual plot of land to start his administrative work of building construction. 
\textit{Living City}, as the name suggests, is a full real-time online game: this means, mainly, that each in-game action implies a real time request, expressed in real minutes, hours or even days, depending on the degree of complexity of the task. 
Starting from the lowest levels, i.e. \textit{village}, \textit{town}, \textit{small city}, and so on, up to \textit{metropolis}, the level of the plot of land grows unlocking more and more new options and tasks over the time, through eight levels of increasing difficulty. 

The system to develop buildings is complex, so we schematized it as follows: there are seven categories of buildings, each category has got several sub-categories.
Each player can build at the same time only one structure for each category, so the system encourages users to follow a planning strategy to maximize development. 
This is the classification of buildings in \textit{Living City}:

\begin{itemize}
  \item Free Areas (19 total buildings), i.e. Residential, Commercial, Industrial, etc;
	\item Transport (18 total buildings), i.e. Transportation, Public places, etc;
	\item Services (16 total buildings), i.e. Power and Water supply, Garbage collection, etc;
	\item Institutional Buildings (17 total buildings), i.e. Institutions, Police, Firefighters, Army, etc;
	\item Public Facilities (19 total buildings), i.e. Education, Health-care, Worship
	\item Tourism and Others (18 total buildings), i.e. Accommodation, Various
	\item External Buildings (40 total buildings), i.e. Commercial, Industrial, Tourism
\end{itemize}

Each point is self-explanatory except the latter: \textit{external buildings} comprises all the structures which can be placed in plots of land belonging to other players after acquiring a building permit, for instance winning a procurement for providing a particular service. 
\textit{Living City} allows up to 147 total buildings, a \textit{huge} number if compared to other management games which count a couple of dozens of possibilities, some mutually esclusive (e.g. Ogame \footnote{http://www.ogame.org}, Travian \footnote{http://www.travian.com}, etc.). 
Structures require virtual money and actual time to be developed: for each level of development of the city, the player can build a couple of structures for each category. 
Buildings themselves grow by levels, from 1 to 12: each upgrade requires more time and money, usually less than the construction from scratch, and grants some small bonuses. Indeed, reaching the cap level of a building usually gives the most useful bonus relating to the category of the building itself. 
There are also services provided by buildings.
We will cover this aspect in next sub-section.

\subsubsection{Management}
The management macro-area is the core of the simulation: players have to tackle economic problems and a wise management of funds is central. 
There are many notable sub-sections: \textit{resources management} allows the player to establish how much money allocate to specific services provided to virtual citizens (usually called \textit{sims}), to allocate funds to services maintenance, to tax collection, etc. 
All these parameters (in addition to many others) are reflected on the virtual \textit{quality of life} in the city, influencing aspects like population distribution, law enforcement, welfare, health-care, employment, pollution and so on. 

The degree of simulation is high and this implies that users should take advantage of some powerful tools like the \textit{economic framework} and the \textit{projections instrument}: the former summarizes all economic aspects of simulation, including income (i.e. income taxes, earnings from provided services, etc.) and costs sustained (i.e. utilities and services costs, etc.), payable and receivable interest, and so on, of the current and the past week. 
The latter is a dynamic real-time overview of the statistical data concerning the virtual quality of life. 
Each decision (e.g. modifications of configurations of resources, in-game actions, etc.) taken by the Mayor is timely reflected on the data projections, and all values related to the virtual quality of life tend to these projections.
The player can thus get a feeling on the evolution of the game, and try to pursue wise decisions.

There are two other gameplay aspects: licenses and provided services. 
Each plot of land can ensure, depending on the level, a growing number of free slots; these can be licensed to other Mayors, in which can be built external buildings, useful to provide services required by the city. 
The development of such structures allows to provide specific services related to the category of the building, so there are services related to utilities, supplies (i.e. power and water supply), transportation, tourism, etc. 
Players acquire building rights for external slots by winning a procurement. 
Likewise, the same players, by issuing calls for procurements, allow others to provide them services: this system ensures a reciprocal benefit to cities active in the licenses-procurements market. 
There are sections with tools useful to monitor procurements, both for issuing a call for new ones and to compete in existing procurements called by other Mayors. 
This play aspect is the core of the collaborative and cooperative simulation and is maintained by three mechanisms (i.e., bid auctions, direct offers and fixed cost offers) to ensure that players come to an agreement in different forms of transaction.

\subsubsection{Tools, Community and Game Guide}
The last macro-area comprises three sections: tools, community and game guide. 
We are not focusing on the former as it presents standard elements, like a \textit{search tool}, a \textit{news page}, a \textit{messaging system}, etc. 

Much more innovative is the \textit{community section}, including some key aspects of the social network phenomenon: each player can create a personal profile including information, share elements with others and so on, contributing in creating a solid community of users. 
Players can also rely on the in-game forum, embedded in the same platform, which permits the creation of links between discussion threads, personal and game profiles. 
The forum, for instance, allows players to integrate projections, statistics, graphs, etc. in their posts, to share game information, advices and tips. 
The \textit{community section} presents pages for top rankings, divided by categories of ranking (fame, virtual quality of life, richness, number of buildings, etc.).
Another innovative aspect introduced in \textit{Living City} is the embedded game guide: thanks to the Web-based nature of the game, we developed guide pages linked to game sections, and created \textit{macros}, coded in JavaScript, that perform useful tasks for players (e.g. players can be warned when a building task is completed by a pop-up message, etc.). 
The main idea was inspired by the work of Kletsch and Volk \cite{Kletsch:ajaxengine} who integrated AJAX \cite{Garrett:ajax} in the engine of a BBMMOG.

\subsection{Game Balance}
The key to develop an enjoying game is \textit{balance}: an important amount of time during the \textit{Living City} design and testing phases was spent studying and fixing formulas and algorithms which manage consumable entities, like money, and real-time aspects, like the passing of time. 
Players' feedback were fundamental to understand problems and to fix them, in particular those related to the impossibility of executing some actions while respecting economic and temporal limits imposed by some unbalanced functions. 
One of the most interesting challenges we faced was to balance the time a player should spend to proceed in the game: this task is not trivial because we had to find solutions that do not favor avid players who were going to spend hours playing, but, at the same time, that reward long time playing users, benefiting the \textit{long planning} game style instead of the \textit{addiction} game style. 
At the same time, balancing money earning and, generally, flow of money was pursued. 

\subsubsection{Time Balancing}
We studied the real-time intervals required to build structures, with a recursive exponential function in order to compute the amount of required time needed for each stage of construction and improvement of buildings: 
\begin{displaymath}
	RequiredTimeAtStage(x) = RequiredTimeAtStage(x-1)^\alpha
\end{displaymath}
where we empirically assumed $1.04 \le \alpha \le 1.06$ and \textit{RequiredTimeAtStage(1)} being the required time to build from scratch.

We carefully compiled a table of all required \textit{build from scratch} times for the 147 structures included in \textit{Living City}, ranging from 90 to 3000 seconds, depending on the complexity of the building and the level required to build it. We obtained 29 different required times, grouped in 5 \textit{time classes} (referred to buildings with similar required times to  \textit{build from scratch}) reported in Table 1.

\noindent
\begin{table}[!h]
  \begin{tabular}{|l|c|l|}
	\hline
	Class 	&	Required Time 							&	$\alpha$\\
	\hline
	\hline
	I	&	90 $\le$ \textit{RequiredTimeAtStage(1)} 	$<$ 300		&	1.06\\
	II	&	300	$\le$ \textit{RequiredTimeAtStage(1)} 	$<$ 600		&	1.055\\
	III	&	600	$\le$ \textit{RequiredTimeAtStage(1)} 	$<$ 1200	&	1.05\\
	IV	&	1200 $\le$ \textit{RequiredTimeAtStage(1)} 	$<$ 1800	&	1.045\\
	V	&	\textit{RequiredTimeAtStage(1)} 		$\ge$ 1800	&	1.04\\
	\hline
  \end{tabular}
  \label{tab:table1}
  \caption{Time Classes and Required Times}
\end{table}
The value of $\alpha$ is in inverse proportionality to time classes, since we realized that it was really hard (to nearly impossible) to complete all the 12 stages of buildings with high \textit{RequiredTimeAtStage(1)}, because of the exponential aspect.

\begin{figure}[!h]%
	\includegraphics[width=240pt]{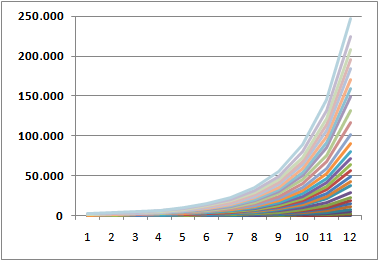}%
	\caption{RequiredTimeAtStage(x) as a function of stages}%
	\label{graph1}%
\end{figure}

In Figure \ref{graph1} the required time is expressed as a function of each of the 12 stages of the building development.
Each line represents one of the 29 \textit{RequiredTimeAtStage(1)} values, and grows exponentially in relation to the development stage of buildings. 
This algorithm defines a smooth exponential function and appears to be the solution which fits  our needs, in term of time balancing. 

\begin{figure}[!h]%
	\includegraphics[width=240pt]{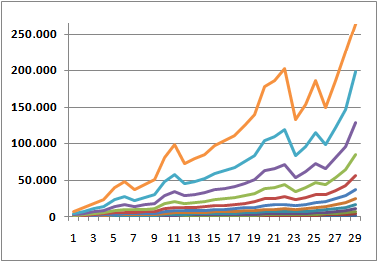}%
	\caption{RequiredTimeAtStage(x) as a function of building completion times}%
	\label{graph2}%
\end{figure}

It is also interesting to analyze the reverse plot: putting the 29 \textit{RequiredTimeAtStage(1)} values on the abscissa, the 12 lines represent the development stages of buildings (Figure \ref{graph2}). 
We observe that lines are not smooth, rather we have an exponential step function, each of the 4 steps reflecting how the variation of $\alpha$ over $[1.04, 1.06]$ affects the function, as a consequence of the transition among time classes. 
The direct consequence of this phenomenon is that, some buildings having a low \textit{RequiredTimeAtStage(1)} in an higher \textit{time class}, at advanced stages of development require less time than those on lower \textit{time class} but with high \textit{RequiredTimeAtStage(1)}, within the time class itself.
This is so because of the inverse proportionality of $\alpha$ relative to growing time classes.
For example, a building requiring 540 seconds ($\alpha=1.055$) for completing stage 1 of construction, will take more than 80000 seconds for upgrading from stage 11 to 12.
On the opposite, a building requiring 600 seconds ($\alpha=1.05$) takes less than 60000 seconds for the same upgrade.

It is also interesting to compute the lower bound for constructing and improving all buildings  in the game up to the cap level: 

\begin{align*}
 MinTime &= \sum_{k=1}^{147} \sum_{x=1}^{12}{RequiredTimeAtStage(x_k)} \\
	 &= 27978980,96 {\rm{\,sec.}} = 7771,93 {\rm{\,hours}} \simeq 324 {\rm{\,days}}.
\end{align*}
where $k$ denotes the $k-th$ building.

\subsubsection{Budget Balancing}
After fixing the function for time balancing, we studied a similar function to balance cash flows in building construction and improvement process:
\begin{displaymath}
BuildingCostAtStage(x) = \beta \cdot BuildingCostAtStage(x-1)
\end{displaymath}
where we empirically assumed $0.10 \le \beta \le 0.15$ and \textit{BuildingCostAtStage(1)} being the required money to build from scratch.

This is a linear-growth function that has been found to be more realistic. 
We established all the 147 \textit{build from scratch} required parameters, in the 2,000.00 to 387,500.00 Euro range, relating to the degree of complexity and the required level of each building. 
We thus obtained exactly 100 different costs; we preferred not to create \textit{cost classes}, rather we calculated the $\beta$ value, once again in inverse proportionality to the cost, normalizing the interval [2,000.00, 387,500.00] to the interval [0.10, 0.15]:
\begin{displaymath}
	\beta = 0.15 - \frac{BuildingCostAtStage(1)\cdot(0.15-0.10)}{(387,500.00-2,000.00)}
\end{displaymath}

This function fits our needs: each upgrading stage costs much less than building from scratch. 
On the opposite the full process of improvement from stage 2 to 12 costs more than the \textit{BuildingCostAtStage(1)}.

$$S = \sum_{x=2}^{12}{BuildingCostAtStage(x)}$$

It is easy to show that, with $\beta \in$ [0.10, 0.15], we have the following boundaries:

$$\frac{3}{2} < \frac{S}{BuildingCostAtStage(1)} < 4$$

\begin{figure}[!h]%
	\includegraphics[width=240pt]{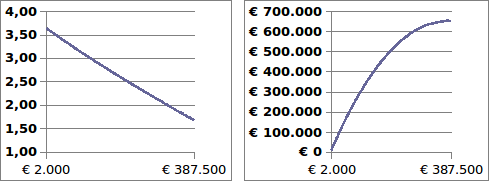}%
	\caption{$\frac{S}{BuildingCostAtStage(1)}$ and S functions}%
	\label{graph3}%
\end{figure}
As can be seen in the left plot represented in Figure \ref{graph3} the $\frac{S}{BuildingCostAtStage(1)}$ function decreases in inverse proportionality to the growth of \textit{BuildingCostAtStage(1)} costs.
We have a linear decrease of money required for improving buildings.
On the right plot in Figure \ref{graph3}, S is plotted as a function of costs: as can be seen it reflects the expected summation function behaviour and also here is possible to see how values are contained in the interval $]\frac{3}{2}, 4[$.
At the boundaries we have:

$$f(2000) = 2000 \cdot 3.64 = 7280$$
$$f(387500) = 387500 \cdot 1.68 = 651000$$

\begin{figure}[!h]%
	\includegraphics[width=240pt]{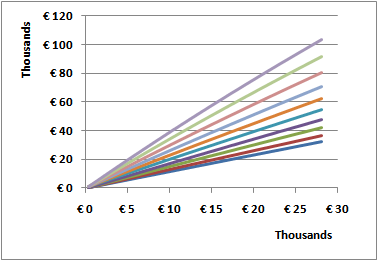}%
	\caption{BuildingCostAtStage(3\dots12) as a function of BuildingCostAtStage(2)}%
	\label{graph4}%
\end{figure}

The linear behaviour of the \textit{BuildingCostAtStage(x)} function is more clearly shown in Figure \ref{graph4} where lines represent the \textit{BuildingCostAtStage(3 \dots 12)}, related to the BuildingCostAtStage(2), represented on the abscissa: it can be seen that the ratio of the two axes is comprised in the $]\frac{3}{2}, 4[$ interval.

\section{The underlying business logic}
MMOGs are fundamentally complex client-server applications; BBMMOGs, in particular, exploit the three-tier architecture \cite{Eckerson:Three} as a viable design pattern which fits well architectural requirements. 
We focus on the business logic tier, exploring some interesting aspects of the logic of \textit{Living City}, e.g. the simulation of time passing (earlier explained), the avoidance of cheating, the bug exploiting in the game and the problem of handling concurrency.

Differently from classic MMOGs, where there are some servers and a client that needs to be installed, in BBMMOGs there is a sort of inversion of the point of view. 
In the former, each client is connected to a central single world, to which, for instance, also administrators could connect and monitor players; in the latter, the instance of the world is replicated in the Web browser of each player, which makes a central control checking impossible, and introduces also updating and persistence problems.


\subsection{Cheating Avoidance}
In accord to Yan and Randell's classification \cite{Yan:cheatingclassification}, we applied cheating avoidance criteria on a large set of possible vulnerabilities and failures. 
All solutions have been designed to ensure an high standard of game fairness as a consequence.
We implemented a series of JavaScript functions, that connect the user interface to the logic layer.
At DBMS level, referential integrity constraints and checks have also been implemented.

Some cheating typologies do not apply to \textit{Living City} because of its intrinsic nature. 
This is the case of exploiting misplaced trust and modifying client infrastructure. 
It can not be applied since there is no any client executable program, replaced by the Web user interface. 
Some others are excluded because of the design of \textit{Living City}, e.g. exploiting machine intelligence, as there are no objects governed by an AI engine, all the playing characters are human. 
Also, cheating related to internal misuse is avoided as administrators cannot own game accounts. 

Apart from Social-Engineering cheating, whose purpose is to steal information, e.g. account user-name and password of other players, which are pretty hard to prevent, we focused on some possible vulnerabilities of the system, i.e., cheating by abusing the game procedures and exploiting bugs.
Two notable examples are illustrated in some detail in the following sub-sections.

\subsubsection{SQL Injection Avoidance}
SQL Injection \cite{Anley:sqlinjection} is a well-known class of techniques, frequently used in BBMMOGs (and, generally, in Web applications), exploiting system vulnerabilities of lack of control and filtering of the input: its aim is the insertion of destructive or spoiling SQL commands through a smart tampering of natural SQL commands sent from the client (the Web user interface), thus directly compromising the database layer integrity. 
\textit{Living City} has been designed with shrewdness: all information that could be retrieved from the game interfaces come from database \textit{views}, i.e., virtual data structures built as a result of queries on tables, multiple table join, aggregating data, etc., with the aim of hiding structural information to users. 
It could be argued that could be not enough: despite views, it could be still possible to attack the database layer through \textit{blind SQL injection techniques} \cite{spett:sqlinjection}, flooding the system with automatic generated SQL commands that attempt to find a vulnerability also without any response from the system itself. 
These blind techniques are potentially dangerous as targeted SQL attacks, so a double security check was designed. 
All SQL commands implemented in the game are parameterized statements \cite{1480579}. So it is possible to change only values of parameters but not structure of statement, thus ensuring an high standard of security, since all parameters are processed by an SQL engine designed specifically to prevent these kind of tampering.
Moreover, all incoming SQL queries from dangerous entry-points (i.e. free text-boxes) are pre-processed looking for destructive or spoilage commands (i.e. drop tables, selecting from master database, altering tables, modifying users, etc.), preventing, de facto, that information could be extracted or deleted through tampering queries. 

\subsubsection{Simultaneously Buildings Avoidance}
Another important difference among standard MMOGs and BBMMOGs is related to \textit{inner concurrency}: obviously, a player may not open two clients and log in the same world with the same account in a well designed MMOG. 
Vice versa, in a BBMMOG players could execute multiple operations by simply opening multiple browser tabs. 
Indeed, a conceptual gameplay limit of \textit{Living City} is that a player cannot simultaneously build two structures belonging to the same category: without any check it could be possible for a user to open multiple tabs on the same category page buildings, thus starting different, albeit theoretically mutually-exclusive, tasks. 

In order to avoid this type of cheating, we implemented a double check that uses AJAX, business logic and database constraints. 
It works as follows.
Each time a player executes a building task, a JavaScript function asynchronously calls a check procedure aimed at verifying the value of a Boolean static variable related to the building category: a \textit{true} value means that another building task of the same category is currently running and this information is notified to the player. 
It could happen that a skilled attacker could launch multiple tasks, from multiple pages, exactly at the same moment, so we also designed integrity constraints directly at the database level.
For instance, the timestamp of each building task is checked in order to verify the legitimacy of any other task in the same category during the same game session. 

\subsection{Handling Concurrency}
A similar approach is used to handle \textit{outer concurrency}: there are elements in the game, like fixed price procurements, which could be simultaneously awarded by two or more players (or, by an extreme, by the same player from different instances of the Web browser). 
Also the recruitment step is very sensitive: unemployed staff members (regardless of whether they have ever been employed or they have been fired or their Mayor left the game and his account was deleted by the system) can be hired by Mayors searching a \textit{common marketplace}; all these cooperative/competitive gameplay situations require an accurate handling of concurrency. 
We called these phenomena \textit{outer concurrency} because of the interaction among different players. 
As a consequence of its intrinsic nature of Web application, in \textit{Living City} all information that must be shared among players, are \textit{published} on HTML pages, which are dynamically generated at the time of the user request.
This implies that data could become obsolete while a player is reading them, since players in real-time concur for the same resources. 

We approached and solved this problem by exploiting AJAX asynchronous updating of pages. Database locking features at the DBMS tier was also adopted, in order to save the referential integrity when AJAX is not enough. 
This situation causes a relevant amount of transactions that are to be executed in a relatively short time interval.
In order to assure good performances, we carefully optimized all the real-time checking algorithms which handle the concurrency of actions. 
Fortunately all these operations are relatively simple to implement, since the only variable to be checked is the availability of the requested resource.

\section{Results and final remarks}
The beta-testing period lasted almost a year: during this time, only a couple of players reached the highest ranking in all possible tasks and developed all structures, thus completing the game. 
Their achievement confirms that at least 11 months of gaming are needed to reach the cap level of \textit{Living City}: indeed most players reached medium-high levels with 6-8 months of moderate playing. 
A good number of players demonstrated a notable management capability and a profitable ability to maximize tasks and activities. 
For instance, it is not a trivial task to optimize the development time of buildings, because players can build only one structure for each category at the same time. 
Interestingly, analogous considerations are found in Allegra et al. \cite{Allegra:environments} work on learning.
  
\subsection{User activity Analysis}
We studied results analyzing log data related to 11 months (from January 7th to December 7th, 2009) of activity.
During this period \textit{Living City} had 9775 registered and active players and almost 165000  visits; this means that the game had 6\% average conversion of visitors into new players each day. Moreover 29\% of the traffic each day came from new visitors (i.e., 71\% of returning visitors are registered players). 
The server displayed over 3 million pages to players, with an average of 16.7 pages per visit. 
We estimated that the number of average visits per day from the same player was 3.06, with an average duration of 10.15 minutes per visit. 
We also have statistics about the game activity:
\begin{itemize}
	\item 192310 buildings (avg. 19.67 building/player)
	\item 75627 calls for procurements (avg. 7.74/player)
	\item 8919 offers for procurements (avg. 0.9/player)
	\item 11488 services provided (avg. 1.17/player)
	\item 120722 staff members employed (avg. 12.35/player)
	\item 7327 capital investments (avg. 0.75/player)
\end{itemize}

We also gathered some data about social activities: players exchanged 2662 personal messages during this period and discussed a lot on the in-game forum; they created 689 threads, with 3363 total posts (almost 5 posts/thread on average). 
The percentage of threads without any reply is less than 6\%. 

\subsubsection{Considerations}
\begin{figure}[!h]%
	\includegraphics[width=240pt]{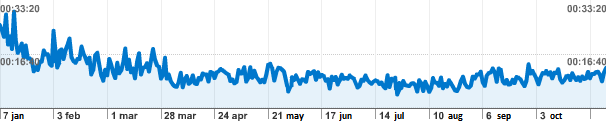}%
	\caption{Average playing time}%
	\label{average-time}%
\end{figure}

It is interesting to analyze the time spent playing the game. 
In Figure \ref{average-time} we plotted the average duration per visit.
A similar behaviour should result from the plot of the \textit{history} of the time spent on the game by a single player. 
The start-up of the game requires much time per visit (almost half an hour for each game session), as operations and tasks are more frequent and last less time. 
Higher levels require 10-15 minutes-per-game sessions. 
Considering that each player in average starts 3 game sessions each day, the time spent is less than two hours a day: the game makes players involved but should not cause addiction. 

\begin{figure}[!h]%
	\includegraphics[width=240pt]{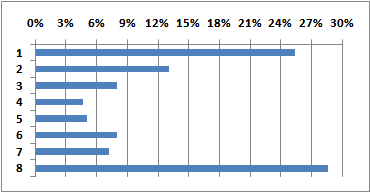}%
	\caption{Player distribution over levels}%
	\label{level-distribution}%
\end{figure}

Figure \ref{level-distribution} shows the distribution of time percentage for each level of the game.
There is an important contribution from new players.
Physiologically, some of them will leave the game in the future.
At the same time, almost a third of the users who started to play reached the cap level (indeed many features are unlocked at level eight, so the game is anything but completed).

Analyzing in-game activities, we concluded that we must improve some features, like the procurement and the investment systems: although these gameplay aspects are massively introduced at later stages game, only a relatively small number of players took advantage of them (in fact the average values of this time are not really significant, because the most important number of procurement offers and capital investment came from only a couple hundreds of players). 
\subsection{Social Interaction}

\subsubsection{Social And Psychological Aspects}
Klimmt et al. \cite{Klimmt:Exploring} studies opened a new perspective on the MMOGs design: they found that \textit{'competition, in contrast, seems to be less important for browser gamers than for users of other game types'}.
This result is pretty important as it introduces an unexpected orientation to cooperative gameplay which is usually nonexistent in MMOFPS and low in MMORPGs.
Massive collaborative tasks have been designed to be the core of evolution in the game, thus making impossible to reach the cap level of the game playing alone. 
Our experience with \textit{Living City} confirms that in BBMMOG players like cooperation, and do not dislike collaboration together with competition to tackle game challenges (e.g. concurring for fame and climbing rankings).
Yee \cite{Yee:Motivations} submitted a list of questions about motivations of playing MMOGs to a sample of 3000  regular players, and discovered how aspects of achievement, social and immersion components interact and determine the in-game behaviour of players. 
Socializing, relationship-building and team-working appears to be relevant and as important as other achievements, usually considered more influencing, like advancing in the game, role-playing and competition. 
We can say that both Klimmt's and Yee's expectations and results are empirically confirmed through the approval by players of gameplay and collaborative dynamics of \textit{Living City}. 


%
\bibliographystyle{abbrv}
\bibliography{sigproc}  
%

\end{document}